\documentclass[11pt]{article}
\usepackage{amsmath,amsthm,amssymb,amscd}
\usepackage{hyperref}
\usepackage{geometry}
\usepackage{tikz-cd}
\usepackage{booktabs} 
\usepackage{graphicx} 
\usepackage{multirow}
\usepackage{tikz}
\usetikzlibrary{calc,positioning,3d}
\usepackage[normalem]{ulem} 
\usepackage{imakeidx}
\usepackage{makecell}
\usepackage{diagbox}
\usepackage{xr-hyper}

\usepackage{amsmath,amsthm,amssymb,amscd}
\usepackage{hyperref}
\usepackage{geometry}
\usepackage{tikz-cd}
\usepackage{booktabs} 
\usepackage{graphicx} 
\usepackage{multirow}
\usepackage{tikz}
\usetikzlibrary{calc,positioning,3d}
\usepackage[normalem]{ulem} 
\usepackage{imakeidx}
\usepackage{enumitem}
\makeindex

\usepackage{subfigure} 
\usepackage{subcaption}
\usepackage{caption}
\usepackage{booktabs}
\usepackage{longtable}
\usepackage{array}\usepackage[all]{xy}

\usepackage{makecell} 
\makeindex

\usepackage{subfigure} 

\newtheorem{theorem}{Theorem}[section]

\theoremstyle{definition}
\newtheorem{definition}[theorem]{Definition}

\DeclareMathOperator{\im}{im}

\DeclareMathOperator{\Lk}{Lk}

\DeclareMathOperator{\Spec}{\mathbf{Spec}}

\makeatletter
\newcommand*{\addFileDependency}[1]{
	\typeout{(#1)}
	\@addtofilelist{#1}
	\IfFileExists{#1}{}{\typeout{No file #1.}}
}

\makeatother
\newcommand*{\myexternaldocument}[1]{%
	\externaldocument{#1}%
	\addFileDependency{#1.tex}%
	\addFileDependency{#1.aux}%
}

\myexternaldocument{SI}

\title{Persistent local Laplacian prediction of protein-ligand binding affinities}
		
	\author{Jian Liu$^1$ and
	Hongsong Feng$^{2}$ \footnote{Corresponding author: Hongsong Feng (hfeng2@charlotte.edu).~},\\		
	 $^1$ Mathematical Science Research Center, \\ 
	 	Chongqing University of Technology, Chongqing 400054, P. R. China\\
		$^2$ Department of Mathematics and Statistics,\\
		University of North Carolina at Charlotte, Charlotte, NC 28223, USA\\
	}

	\date{\today}
\begin{document}
	\maketitle 
\begin{abstract}	    
Accurate prediction of protein–ligand binding affinity remains a central challenge in structure-based drug discovery. The effectiveness of machine learning models critically depends on the quality of molecular descriptors, for which advanced mathematical frameworks provide powerful tools. In this work, we employ a novel mathematical theory, termed the persistent local Laplacian (PLL), to construct molecular descriptors that capture localized geometric and topological features of biomolecular structures. The PLL framework addresses key limitations of traditional topological data analysis methods, such as persistent homology and the persistent Laplacian, which are often insensitive to local structural variations, while maintaining high computational efficiency. The resulting molecular descriptors are integrated with advanced machine learning algorithms to develop accurate predictive models for protein–ligand binding affinity. The proposed models are systematically evaluated on three well-established benchmark datasets, demonstrating consistently strong and competitive predictive performance. Computational results show that the PLL-based models outperform existing approaches, highlighting their potential as a powerful tool for drug discovery, protein engineering, and broader applications in science and engineering.
 \end{abstract}
	
	
{\setcounter{tocdepth}{4} \tableofcontents}
	 \setcounter{page}{1}
	 \newpage	
	
\section{Introduction}

Accurate prediction of protein–ligand binding affinity is central to drug design and discovery, as it dictates how small-molecule therapeutics recognize and modulate their biological targets and therapeutic efficacy. Despite its importance, drug discovery remains time-consuming, costly, and characterized by high failure rates \cite{fleming2018computer}. Traditional computational approaches, including molecular docking \cite{lyu2019ultra,kitchen2004docking,pinzi2019molecular,pagadala2017software}, free energy perturbation \cite{wang2015accurate}, and empirical scoring functions \cite{sliwoski2014computational}, have contributed significantly to structure-based drug design. However, they often suffer from limited accuracy, high computational cost, and difficulty in capturing complex interaction mechanisms, particularly in large-scale virtual screening and the exploration of novel chemical spaces.

Recent advances in machine learning (ML) and artificial intelligence (AI) have transformed drug discovery by enabling data-driven modeling of complex molecular systems \cite{jumper2021highly,baek2021accurate,lin2023evolutionary,song2024multiobjective}. ML models have achieved remarkable success in predicting protein structures, protein–ligand binding affinities, and other key molecular properties by learning nonlinear relationships directly from experimental data \cite{li2015improving,feinberg2018potentialnet}. In AI-assisted drug design, molecular descriptors play a critical role in model performance and are widely used in quantitative structure–activity/property relationship analyses and virtual screening pipelines \cite{ballester2010machine,wang2017improving,pan2022aa}. Because the biological functions of macromolecules depend strongly on their complex three-dimensional (3D) structures and spatial interactions, substantial effort has been devoted to developing 3D structure-based molecular descriptors. Although these descriptors, together with advanced deep learning architectures—such as graph neural networks, convolutional neural networks, and transformer-based models—have improved predictive accuracy \cite{nguyen2021graphdta,wallach2015atomnet,ragoza2017protein}, significant challenges remain in effectively capturing essential physicochemical information such as hydrogen bonding, electrostatics, and van der Waals forces, while maintaining a simplified representation of molecular structures.

To address these challenges, advanced mathematical methods grounded in algebraic topology, geometry, and spectral theory have emerged as powerful tools for biomolecular representation \cite{nguyen2020review,meng2021persistent,liu2023persistent,wang2020persistent}. In particular, topological data analysis methods, such as persistent homology (PH) \cite{carlsson2009topology,edelsbrunner2008persistent}, provide multiscale abstractions of molecular structures that capture essential geometric and topological features. PH has demonstrated strong performance in protein–ligand binding affinity prediction and achieved top-tier results in international drug design competitions such as the D3R Grand Challenges \cite{nguyen2019mathematical,nguyen2020mathdl}. Beyond the success of PH, persistent combinatorial Laplacian, also known as persistent Laplacian (PL) techniques \cite{wang2020persistent,memoli2022persistent,chen2024multiscale}, offers enhanced molecular representations by encoding additional geometric information through the harmonic and nonharmonic spectra of Laplacian matrices. PL has been successfully applied to tasks including protein engineering \cite{qiu2023persistent} and the accurate prediction of emerging dominant SARS-CoV-2 variants \cite{chen2022persistent}. 

As a further advancement, we recently proposed a novel TDA framework, termed persistent local Laplacians (PLL) \cite{liu2026local}. While conventional PH and PL methods primarily capture global topological structures and are largely insensitive to local data characteristics, PLL explicitly incorporates localization to characterize both local topological features and geometric information in complex datasets. In particular, PLL enables the simultaneous capture of global topology and local geometric–physical interactions in biomolecular systems, which are crucial for describing the physicochemical interactions that govern molecular functions.

The fundamental idea of PLL lies in investigating the topological and geometric invariants within the neighborhood of a specific point in a dataset. According to the excision theorem in algebraic topology, the local homology group is independent of the global structure of the entire space or dataset \cite{hatcher2002algebraic}. This key theoretical property implies two significant advantages: first, the computation of local homology and Laplacian only needs to be performed within the neighborhood of a single point, which drastically reduces the computational complexity. Second, the homology and Laplacian calculations at different points are inherently parallelizable, leading to a substantial improvement in computational efficiency. Another noteworthy advantage of PLL is the scalability of local feature computation. Specifically, even when the size of the entire dataset expands, the precomputed local topological and geometric information remains unaffected and does not require re-calculation. Furthermore, PLL establishes a multi-scale detection approach at the local level, which can be analogized to a microscope with adjustable magnification. This unique characteristic enables us to conduct targeted investigations into the information of specific positions within a target object, thereby providing fine-grained insights into local structural properties that were previously difficult to capture with traditional methods.

The objective of this work is to employ PLL theory to develop highly accurate machine learning models for protein–ligand binding affinity prediction. We evaluate our approach on three widely used benchmark datasets from the PDBbind database—PDBbind-v2007, PDBbind-v2013, and PDBbind-v2016 \cite{liu2015pdb}. Given the complexity of these datasets, which involve intricate three-dimensional (3D) protein–ligand structures and diverse physicochemical interactions, we propose element-specific PLL models to effectively capture the underlying interaction patterns in biomolecular systems. As demonstrated in our results, the proposed method outperforms those published models and achieve state-of-the-art performance across these benchmark datasets.

The rest of this paper is organized as follows. Section~2 is devoted to the results of PLLML for protein–ligand binding. Methods are described in Section~3. This paper ends with a conclusion.
 
\section{Results} 

\begin{figure}
\centering
\includegraphics[width=0.85\textwidth]{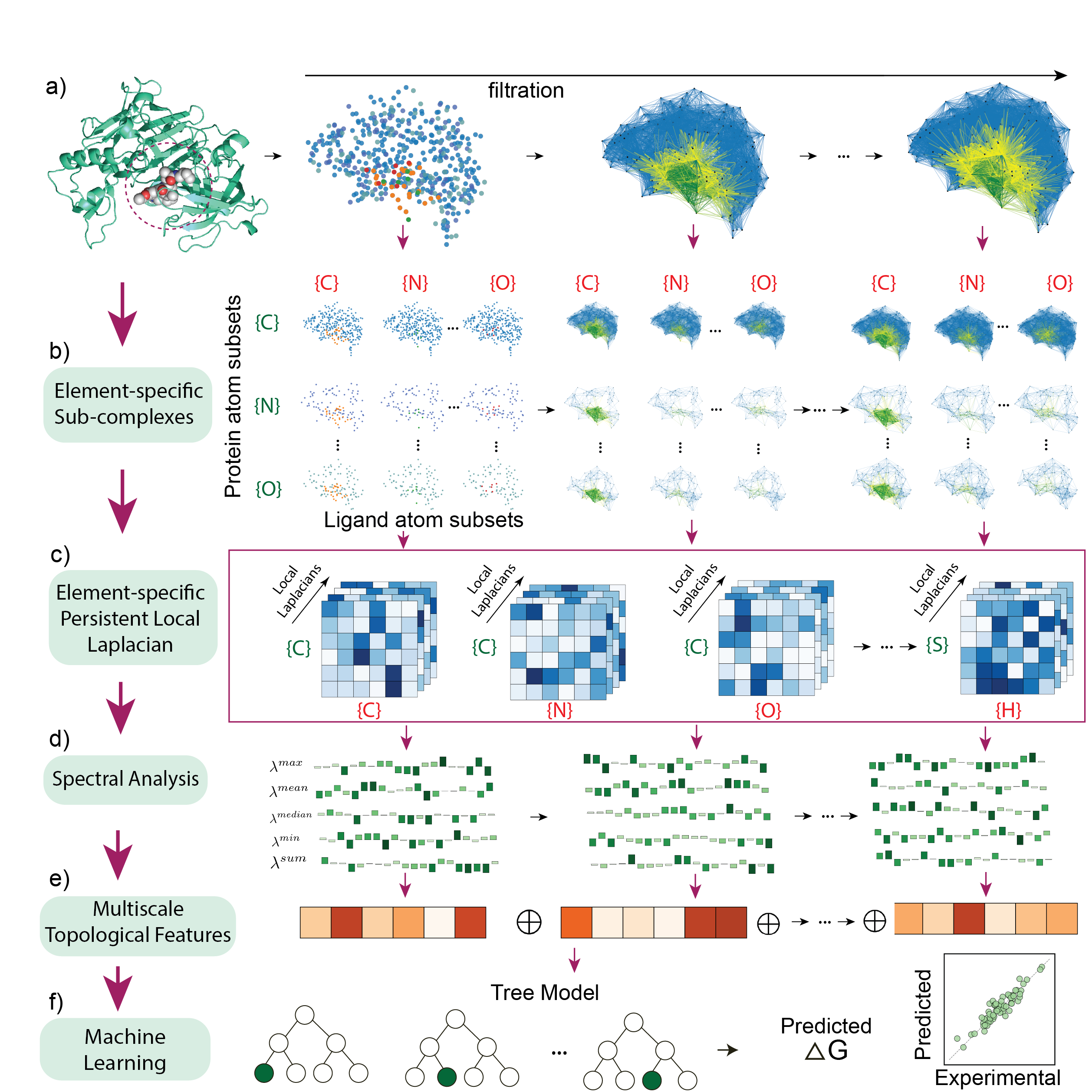}
\caption{Conceptual diagram of the persistent local Laplacian (PLL) platform for protein–ligand binding affinity prediction. a) Construction of a Vietoris-Rips-complex–induced filtration from atoms in a protein–ligand complex. b) Generation of nested families of element-specific subcomplexes along distance-based filtration parameters. c) Formation of atom-specific local Laplacian matrices to characterize atomic interactions within protein–ligand subcomplexes. d) Spectral analysis of the local Laplacian matrices to derive discriminative features for protein–ligand subcomplexes. e) Multiscale topological features for protein–ligand complex are established by concatenating features of protein–ligand subcomplexes. f) Utilization of PLL-based features for downstream binding affinity prediction and analysis using gradient-boosting decision tree models.}
\label{Fig:concepts}
\end{figure}

We evaluate the learning capability of the proposed PLL framework through the task of protein–ligand binding affinity prediction, a central challenge in computational drug discovery. Our experiments utilize the aforementioned three widely used benchmark datasets from PDBbind database~\cite{liu2015pdb}. Each dataset provides experimentally measured binding affinities along with high-quality 3D structures of protein–ligand complexes, making them a gold standard for assessing predictive models in this domain \cite{feng2025mayer, nguyen2019agl, shen2024knot, meng2021persistent, liu2023persistent, cang2017topologynet,feng2025caml}. Summary statistics for dataset sizes and training–test partitions are reported in \autoref{table:datasets-information}.

To characterize the geometric and interaction patterns within each complex, we employ persistent local Laplacian (PLL) theory to construct three-dimensional molecular descriptors tailored to protein–ligand structures. These resulting descriptors are subsequently integrated with gradient boosting decision tree (GBDT) algorithm to build regression learning model. A detailed description of the molecular featurization pipeline and model configuration is provided in the \hyperref[sec:method]{Method} section. The overall pipeline of our machine learning framework is illustrated in \autoref{Fig:concepts}.

We train the models ten times using different random seeds and evaluate their performance using two metrics: the Pearson correlation coefficient ($R$) and the root-mean-square error (RMSE). The predicted binding affinities are compared with the corresponding experimental values. As shown in the first column of \autoref{table:Pearson-values-PDBbind}, the average $R$ values of the PLL-based machine learning models over the ten runs for the three datasets are 0.813, 0.791, and 0.850, respectively. In addition, the PLL model achieves an average $R$ value of 0.818 across the three datasets. These high $R$ values demonstrate the effectiveness and reliability of the PLL-based approach for protein–ligand binding affinity prediction. The low RMSE values (in kcal/mol) further confirm the strong predictive performance of our models.

\begin{table}[htb!]
	\small
	\centering
	\begin{tabular}{c  | c | c |c}
		\hline
		\textbf{Dataset} &  \textbf{PLL} & \textbf{Transformer} & \textbf{PLL+ Transformer} \\
		\hline
		PDBbind-v2007 & 0.813(1.971)& 0.785(2.006) & \textbf{0.827 (1.925)} \\
		PDBbind-v2013 & 0.792(1.976)& 0.791(1.974) & \textbf{0.813 (1.932)} \\
		PDBbind-v2016 & 0.850(1.672)& 0.836(1.713) & \textbf{0.861 (1.646)} \\
		\hline
		Average & 0.818 (1.873) &0.804 (1.898) &\textbf{0.834 (1.834)}\\
		\hline
	\end{tabular}
	\caption{Modeling performance of different strategies on the test sets of the three PDBbind benchmark datasets. The evaluation metrics are the Pearson correlation coefficient ($R$) and the root-mean-square error (RMSE, in kcal/mol, in the parenthesis). Each model was trained and evaluated over ten independent runs using different random seeds, and the reported results are the averaged metric values. PLL denotes the persistent local Laplacian-based machine learning models. Transformer refers to sequence-based models using natural language processing. PLL + Transformer represents the consensus predictions obtained by averaging the outputs of the PLL and Transformer models, which is used as our final model referred to as PLLML.
	}
	\label{table:Pearson-values-PDBbind}
\end{table}

To further enhance predictive performance, we incorporate natural language processing (NLP)–based molecular features to develop an additional set of machine learning models. Pretrained NLP models generate molecular representations directly from molecular sequences. Specifically, we extract molecular features from protein amino acid sequences using a protein language model \cite{rives2021biological} and from small-molecule SMILES strings using a small-molecule language model \cite{chen2021extracting}. These sequence-derived features are concatenated and combined with the gradient boosting decision tree (GBDT) algorithm to build Transformer-based models, whose performance is reported in the second column of \autoref{table:Pearson-values-PDBbind}. On average, the PLL-based models achieve higher Pearson $R$ values than the Transformer-based models.

Compared to the sequence-based models, PLL models capture richer biophysical information encoded in 3D structural data. This is achieved by effectively addressing structural complexities, high dimensionality, and the multiscale and multiphysical interactions in molecular 3D structures. Superior performance of PLL models are obtained over the Transformer models, as shown in \autoref{table:Pearson-values-PDBbind}.

Both sequence-based and 3D structure–based models offer distinct strengths, and we leverage their complementary advantages to further enhance predictive performance. To this end, we generate new predictions by averaging the outputs of the PLL and Transformer models, resulting in a consensus model referred to as PLLML. The final column of \autoref{table:Pearson-values-PDBbind} reports the performance of this consensus approach. The integrated models have substantially improved accuracy, achieving an average Pearson $R$ value of 0.834 across the three datasets. PLLML’s reliability is further demonstrated by the strong alignment between experimental and predicted binding affinities, as visualized in \autoref{Fig:distributions}. 

\begin{figure}[!htb]
	\centering
	\includegraphics[width=0.8\textwidth]{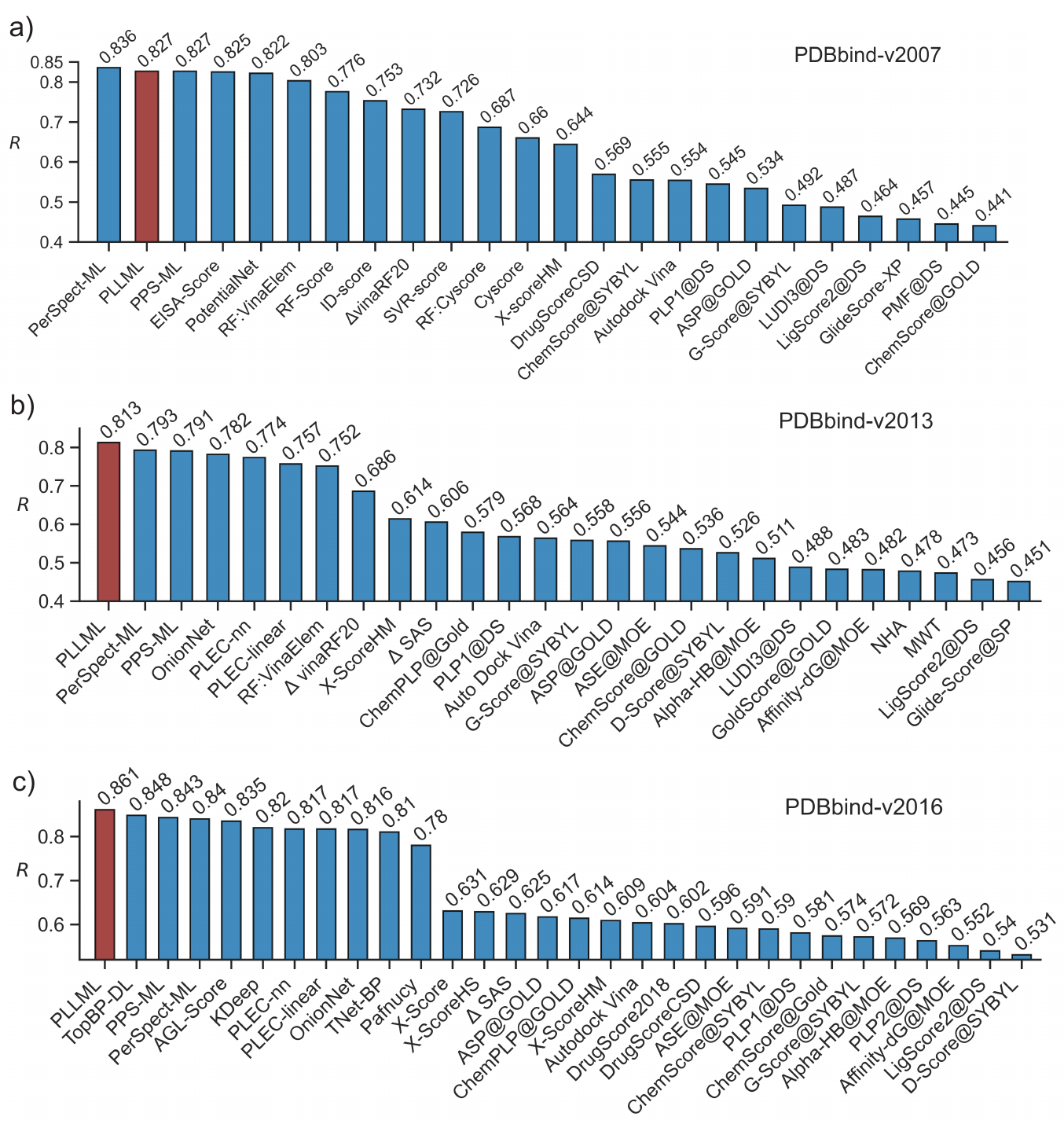}
	\caption{Comparison of the predictive performance of the PLLML model with other published models in terms of the Pearson correlation coefficient ($R$) on three PDBbind datasets.}
	\label{Fig:R-comparison}
\end{figure}

\begin{figure}[!htb]
	\centering
	\includegraphics[width=0.82\textwidth]{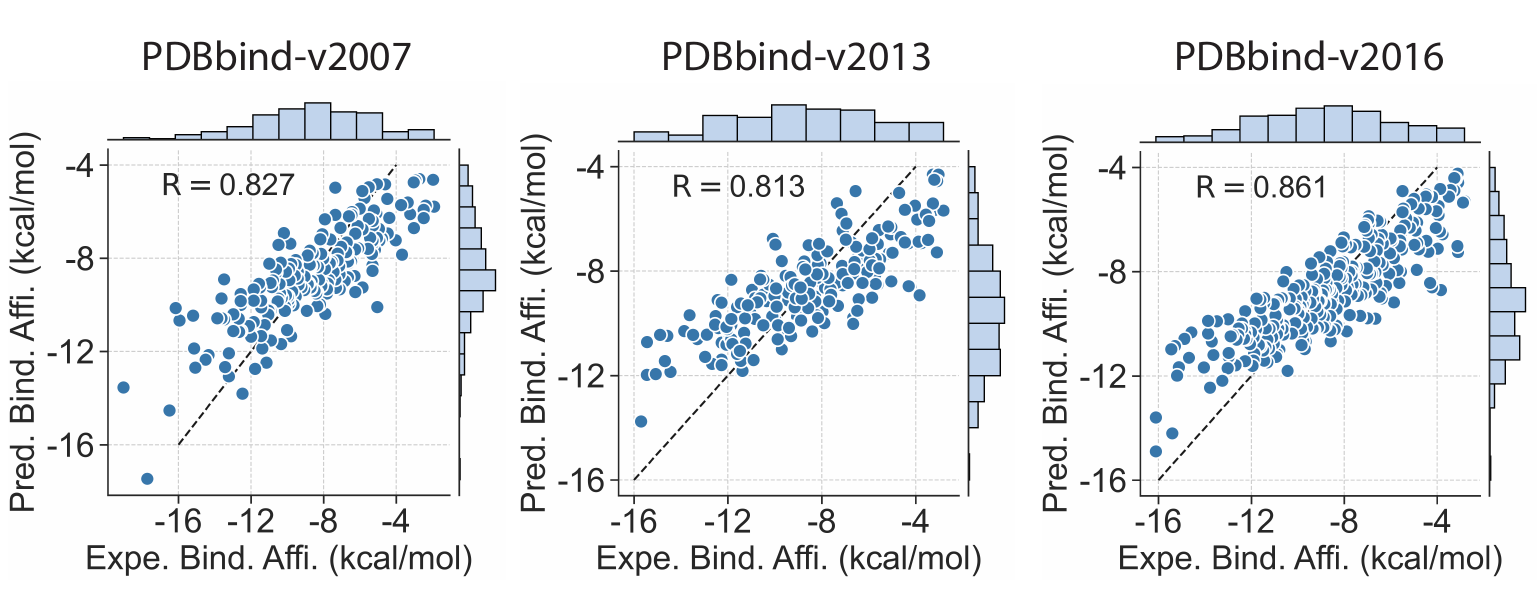}
	\caption{ A comparison between the experimental and predicted binding affinities from our PLLML model on three PDBbind datasets.}
	\label{Fig:distributions}
\end{figure}

Our PLLML models outperform existing state-of-the-art approaches, achieving superior predictive accuracy in binding affinity estimation as shown in \autoref{Fig:R-comparison}. On the PDBbind-v2016, our PLL model achieves R value of 0.850 without the boost from transformer model, which is superior to most published machine learning models. Numerous competitive models grounded in mathematical and physical principles \cite{ballester2010machine,wang2017improving,li2023development}, including those based on persistent homology \cite{cang2018representability} and persistent spectral theories \cite{meng2021persistent,liu2023persistent}, have been developed and consistently rank among the strongest performers in this field (see \autoref{Fig:R-comparison}a). On the PDBbind-v2013 test set, our PLLML model achieved a Pearson correlation coefficient ($R$) of 0.813, outperforming the persistent spectral theory–based models PerSpect-ML ($R=0.793$) \cite{meng2021persistent} and PPS-ML ($R=0.791$) \cite{liu2023persistent}. On the PDBbind-v2016 test set, our PLLML model achieved a Pearson correlation coefficient ($R$) of 0.861, outperforming the persistent homology–based TopBP-DL ($R=0.848$) \cite{cang2018representability} and the persistent spectral theory–based models PerSpect-ML ($R=0.843$) \cite{meng2021persistent} and PPS-ML ($R=0.840$) \cite{liu2023persistent}. Among these published models, PerSpect-ML achieves the best performance, with $R=0.836$, $0.813$, and $0.848$ on PDBbind-v2007, PDBbind-v2013, and PDBbind-v2016, respectively. Although PLLML is slightly inferior to PerSpect-ML on PDBbind-v2007, it yields significantly better predictions on PDBbind-v2013 and PDBbind-v2016. These results highlight the effectiveness of PLLML as a novel analytical tool and demonstrate its potential for developing state-of-the-art models for binding affinity prediction.

\section{Methods} \label{sec:method}

In this section, we will first present the information for the three PDBbind benchmark datasets. Then, we provide an overview of persistent local Laplacian theory (PLL). Next, we describe the molecular featurization based on PLL, followed by natural language processing (NLP) molecular descriptors. Machine learning models are developed by combining these molecular descriptors with advanced machine learning algorithms.

\subsection{Datasets}


The PDBbind database \cite{liu2015pdb} is a widely recognized, curated resource that systematically collects experimentally determined 3D structures of protein–ligand complexes alongside their corresponding binding affinity data, e.g., dissociation constants $K_d$, inhibition constants $K_i$, and Gibbs free energy changes $\Delta G$. We consider three widely used benchmark datasets PDBbind-v2007, PDBbind-v2013, and PDBbind-v2016 to test the performance of our PLL method on protein–ligand binding affinity prediction. The dataset details are shown in \autoref{table:datasets-information}.

\begin{table}[htb!]
	\small
	\centering
	\begin{tabular}{c| c c  c }
		\hline
		\textbf{Dataset} &  \textbf{Total} & \textbf{Training set } & \textbf{Test set}\\
		\hline
		PDBbind-v2007 \cite{cheng2009comparative} & 1300& 1105 &195\\
		PDBbind-v2013 \cite{li2014comparative}&  2959 &2764&195\\
		PDBbind-v2016 \cite{su2018comparative} &  4057 &3767&290 \\
		\hline
	\end{tabular}
	\caption{Details of the datasets utilized for benchmark tests in this study.}
	\label{table:datasets-information}
\end{table}

 \subsection{Persistent local Laplacian theory} 
 
Persistent local Laplacian theory (PLL) \cite{liu2026local} characterizes both local topological structures and geometric information in datasets through the spectral analysis of local Laplacian matrices. This theory is one significant advancement of the two classes of prevalent TDA methods including persistent homology \cite{carlsson2009topology,edelsbrunner2008persistent} and persistent Laplacian \cite{wang2020persistent,memoli2022persistent}. A detailed discussion of the relationships among these methods, as well as the advances and advantages introduced by PLL, is provided in Section~1 of the Supporting Information. Below, we present a brief overview of PLL .
 
 \subsubsection{Simplicial complex}
 Let $d \in \mathbb{N}$ and $\{v_0, v_1, \dots, v_n\}$ be a geometrically independent point set in the Euclidean space $\mathbb{R}^d$. The convex hull
 \[
 \sigma = \left\{ \sum_{i=0}^n \lambda_i v_i \mid \lambda_i \geq 0, \sum_{i=0}^n \lambda_i = 1 \right\}
 \]
 is called an $n$-dimensional simplex (denoted $[v_0,v_1,\dots,v_n]$). In the usual sense, a single point is a 0-simplex, a line segment is a 1-simplex, a triangle (filled) is a 2-simplex, a tetrahedron is a 3-simplex, and so forth. A simplex $\tau$ is a \textit{face} of $\sigma$ if the vertex set of $\tau$ is a subset of that of $\sigma$; a \textit{proper face} is a face with $\dim(\tau) < \dim(\sigma)$. 
 
 A collection $\mathcal{K}$ of simplices in $\mathbb{R}^d$ is defined as a \textit{simplicial complex} if and only if it satisfies two axioms:
 \begin{enumerate}[label=(\arabic*), noitemsep]
 	\item If $\sigma \in \mathcal{K}$, then all faces of $\sigma$ are also in $\mathcal{K}$;
 	\item For any $\sigma_1, \sigma_2 \in \mathcal{K}$, their intersection $\sigma_1 \cap \sigma_2$ is either empty or a common face of both simplices.
 \end{enumerate}
 
 \subsubsection{Simplicial homology}
 For a simplicial complex $\mathcal{K}$, let $C_k(\mathcal{K}; \mathbb{F})$ denote the $k$-th \textit{simplicial chain group} (or chain complex) over a field $\mathbb{F}$, which is the vector space spanned by all $k$-simplices in $\mathcal{K}$. Here, $\mathbb{F}$ is always taken to be the real number field $\mathbb{R}$ or the binary field $\mathbb{Z}/2$. The \textit{boundary operator} $\partial_k: C_k(\mathcal{K}; \mathbb{F}) \to C_{k-1}(\mathcal{K}; \mathbb{F})$ is the linear map
 \[
 \partial_k [v_0, v_1, \dots, v_k] = \sum_{i=0}^k (-1)^i [v_0, \dots, \widehat{v_i}, \dots, v_k],
 \]
 where $\widehat{v_i}$ denotes the omission of vertex $v_i$. One can verify that $\partial_{k-1} \circ \partial_k = 0$, meaning the boundary of a boundary is empty. The $k$-th \textit{simplicial homology group} $H_k(\mathcal{K}; \mathbb{F})$ is defined as the quotient space
 \[
 H_k(\mathcal{K}; \mathbb{F}) = \frac{\ker \partial_k}{\text{im } \partial_{k+1}},
 \]
 where $\ker \partial_k=\{x\in C_k(\mathcal{K}; \mathbb{F})\mid\partial_k x=0\}$ and $\im \partial_{k+1}=\{\partial_{k+1} x\mid x\in C_{k+1}(\mathcal{K}; \mathbb{F})\}$ are $k$-chains that are boundaries of $(k+1)$-chains.
 

\subsubsection{Simplicial Laplacian}

We set $\mathbb{F}=\mathbb{R}$. On the $k$-chain group $C_k(\mathcal{K}; \mathbb{R})$, which is a real vector space spanned by the $k$-simplices of $\mathcal{K}$, we equip the \emph{standard inner product}
\[
\langle \sigma, \tau \rangle =
\begin{cases}
	1 & \text{if } \sigma = \tau, \\
	0 & \text{otherwise},
\end{cases}
\]
for any $k$-simplices $\sigma, \tau$ in $\mathcal{K}$.
With respect to this inner product, the adjoint operator $\partial_{k}^* \colon C_{k-1}(\mathcal{K}; \mathbb{R}) \to C_k(\mathcal{K}; \mathbb{R})$ of $\partial_{k} \colon C_k(\mathcal{K}; \mathbb{R}) \to C_{k-1}(\mathcal{K}; \mathbb{R})$ is defined as
\[
\langle \partial_{k} x, y \rangle = \langle x, \partial_{k}^* y \rangle,\quad \forall x \in C_k(\mathcal{K}; \mathbb{R}), y \in C_{k-1}(\mathcal{K}; \mathbb{R}).
\]
We then have the following adjoint sequence
\begin{equation*}
	\xymatrix{
		C_{k+1}(\mathcal{K}; \mathbb{R})\ar@<0.75ex>[rr]^-{\partial_{k+1}} && C_{k}(\mathcal{K}; \mathbb{R}) \ar@<0.75ex>[rr]^-{\partial_{k}}\ar@<0.75ex>[ll]^-{(\partial_{k+1})^{\ast}} && C_{k-1}(\mathcal{K}; \mathbb{R})\ar@<0.75ex>[ll]^-{(\partial_{k})^{\ast}}.
	}
\end{equation*}
The $k$-th \emph{simplicial Laplacian} (or combinatorial Laplacian) is given by
\[
\Delta_{k} := \partial_{k}^* \partial_{k} + \partial_{k+1} \partial_{k+1}^*,\quad k\geq 0.
\]
In particular, when $\mathcal{K}$ is a 1-dimensional simplicial complex, i.e., an undirected graph, its simplicial Laplacian coincides with the standard graph Laplacian.
  
 \subsubsection{Relative homology}

 Let $\mathcal{K}$ be a simplicial complex and $\mathcal{L} \subseteq \mathcal{K}$ a subcomplex. The chain complex $C_{\ast}(\mathcal{L};\mathbb{R})$ is a subcomplex of $C_{\ast}(\mathcal{K};\mathbb{R})$. For brevity, we write $C_{\ast}(\mathcal{K})$ for $C_{\ast}(\mathcal{K};\mathbb{R})$ (and $C_{\ast}(\mathcal{L})$ for $C_{\ast}(\mathcal{L};\mathbb{R})$) from now on. The \textit{relative chain complex} is defined as the quotient space
 \[
 C_k(\mathcal{K},\mathcal{L}) := C_k(\mathcal{K}) \big/ C_k(\mathcal{L}),
 \]
 which inherits a well-defined boundary operator from $C_\ast(\mathcal{K})$, denoted by
 \[
 \overline{\partial}_k : C_k(\mathcal{K},\mathcal{L}) \to C_{k-1}(\mathcal{K},\mathcal{L}).
 \]
 The homology of this quotient chain complex is called the \textit{relative homology}, denoted $H_k(\mathcal{K},\mathcal{L})$. Relative homology captures $k$-dimensional cycles in $\mathcal{K}$ that become trivial when $\mathcal{L}$ is collapsed or regarded as negligible, and in this way, it isolates the topological features of $\mathcal{K}$ that lie outside $\mathcal{L}$.
 
 \subsubsection{Local homology}
 Let $\mathcal{K}$ be a simplicial complex and $v$ a vertex of $\mathcal{K}$, and let $K$ denote the geometric realization of $\mathcal{K}$.
 \begin{definition}
 	The \textit{local homology} of $\mathcal{K}$ at $v$ in dimension $k$ is defined as
 	\[
 	H_k^{\operatorname{loc}}(\mathcal{K},v)
 	:= H_k\big(K,\, K \setminus \{v\}\big),
 	\]
 	i.e., the $k$-th relative homology group of the pair $(K,\, K \setminus \{v\})$.
 \end{definition}
 Intuitively, $H_k^{\operatorname{loc}}(\mathcal{K},v)$ measures the homological obstruction to removing the vertex $v$. Local homology depends only on an arbitrarily small neighborhood of $v$. Indeed, by the excision theorem, for any open neighborhood $U$ of $v$ in $K$, there is a canonical isomorphism
 \[
 H_k^{\operatorname{loc}}(\mathcal{K},v)
 \cong
 H_k(U,\, U \setminus \{v\}),
 \]
 hence local homology is a purely local topological invariant.
 
 From a data-analytic viewpoint, $H_k^{\operatorname{loc}}(\mathcal{K},v)$ characterizes the intrinsic $k$-dimensional topological structure around $v$ and distinguishes whether $v$ behaves locally like an interior point, a boundary point, or a singular point.

 \subsubsection{Link complex}

To describe the local structure explicitly, we introduce the \textit{link} of a vertex: the link of $v$ in $\mathcal{K}$ is the subcomplex
	\[
	\Lk_{\mathcal{K}}(v)
	=
	\big\{ \tau \in \mathcal{K} \mid v \notin \tau \text{ and } \tau \cup \{v\} \in \mathcal{K} \big\},
	\]
	which records how simplices are attached around $v$ and thus encodes the local connectivity pattern of $\mathcal{K}$ near $v$. A fundamental result relates local homology to the reduced homology of the link
	\[
	H_k^{\operatorname{loc}}(\mathcal{K},v)
	\cong
	\widetilde{H}_{k-1}\big(\Lk_{\mathcal{K}}(v)\big).
	\]
	Thus, all local topological information at $v$ is encoded in the homology of its link.

 \subsubsection{Local Laplacian}
 
 Let $\mathcal{K}$ be a simplicial complex, and let $v$ be a vertex of $\mathcal{K}$. Denote by $\mathcal{L} = \mathcal{K} \setminus \{v\}$ the largest subcomplex of $\mathcal{K}$ that does not contain $v$. Then the relative boundary operator $\overline{\partial}_k: C_k(\mathcal{K},\mathcal{L}) \to C_{k-1}(\mathcal{K},\mathcal{L})$ is referred to as the local boundary operator.
 \begin{definition}
 	The $k$-th local simplicial Laplacian at vertex $v$ is a self-adjoint operator $\Delta_k^{\mathcal{K},v}: C_k(\mathcal{K}, \mathcal{K} \setminus \{v\}) \to C_k(\mathcal{K}, \mathcal{K} \setminus \{v\})$ defined as
 	\[
 	\Delta_k^{\mathcal{K},v} := \overline{\partial}_{k+1} \overline{\partial}_{k+1}^\ast + \overline{\partial}_k^\ast \overline{\partial}_k,
 	\]
 	For $k=0$, this simplifies to $\Delta_0^{\mathcal{K},v} = \overline{\partial}_1 \overline{\partial}_1^\ast$.
 \end{definition}
 
%
%
%

 \subsubsection{Connection to link Laplacian}

 As established earlier, the local homology group of $\mathcal{K}$ at vertex $v$ is canonically isomorphic to the reduced homology of the link complex $\Lk_{\mathcal{K}}(v)$, i.e., $H_k^{\operatorname{loc}}(\mathcal{K},v; \mathbb{R}) \cong \widetilde{H}_{k-1}\left(\Lk_{\mathcal{K}}(v); \mathbb{R}\right)$. This fundamental isomorphism naturally bridges the local Laplacian $\Delta_k^{\mathcal{K},v}$ to the simplicial Laplacian of the link complex $\Lk_{\mathcal{K}}(v)$, which we refer to as the \textit{link Laplacian}.
 
 First, recall that the link complex $\Lk_{\mathcal{K}}(v)$ consists of all simplices $\tau \in \mathcal{K}$ such that $v \notin \tau$ and $\tau \cup \{v\} \in \mathcal{K}$. The link Laplacian $\Delta_{n}^{\Lk_{\mathcal{K}}(v)}$ is defined following the standard simplicial Laplacian: for the $k$-chain group $C_k\left(\Lk_{\mathcal{K}}(v); \mathbb{R}\right)$ equipped with the standard inner product, the link Laplacian is
 \[
 \Delta_{n}^{\Lk_{\mathcal{K}}(v)} := \partial_{k+1}^v (\partial_{k+1}^v)^* + (\partial_k^v)^* \partial_k^v,
 \]
 where $\partial_k^v$ denotes the boundary operator on $\Lk_{\mathcal{K}}(v)$, and $(\partial_k^v)^*$ is its adjoint operator with respect to the standard inner product on $C_k\left(\Lk_{\mathcal{K}}(v); \mathbb{R}\right)$.
 
 The key connection between the local Laplacian $\Delta_k^{\mathcal{K},v}$ and the link Laplacian $\Delta_{k-1}^{\Lk_{\mathcal{K}}(v)}$ lies in their spectral properties, which are directly linked via the isomorphism $H_k^{\operatorname{loc}}(\mathcal{K},v) \cong \widetilde{H}_{k-1}\left(\Lk_{\mathcal{K}}(v)\right)$ and the Hodge-theoretic correspondence between harmonic spaces and homology groups:
 
 \begin{theorem}
 	Let $\mathcal{K}$ be a finite simplicial complex, and $v \in \mathcal{K}$ a vertex. Then there is a unitary equivalence
 	\[
 	\Delta_k^{\mathcal{K},v} \cong  \Delta_{k-1}^{\Lk_{\mathcal{K}}(v)}
 	\]
 	between the local Laplacian and the corresponding link Laplacian at a shifted dimension. In particular, the harmonic space of $\Delta_k^{\mathcal{K},v}$ is isomorphic to the harmonic space of $\Delta_{k-1}^{\Lk_{\mathcal{K}}(v)}$, i.e.,
 	\[
 	\ker \Delta_k^{\mathcal{K},v} \cong \ker \Delta_{k-1}^{\Lk_{\mathcal{K}}(v)}.
 	\]
 \end{theorem}
 In particular, for $k\geq 1$, the local Laplacian can be recovered from the Laplacian of the link. For $k=0$, the local Laplacian is simply scalar multiplication by $\deg v$, the number of edges incident to vertex $v$.

 Denote the spectrum of $\Delta_k^{\mathcal{K},v}$ by $\Spec(\Delta_k^{\mathcal{K},v})$ and the spectrum of $\Delta_{k-1}^{\Lk_{\mathcal{K}}(v)}$ by $\Spec(\Delta_{k-1}^{\Lk_{\mathcal{K}}(v)})$, respectively. The above theorem implies a direct correspondence between the spectral features of the local Laplacian and the link Laplacian.
 
 The link Laplacian provides an effective computational tool for the local Laplacian. Although working with quotient relative complexes is often involved, the link complex admits much simpler computations.
 
 \subsubsection{Persistent Local Laplacian}
 
 We now extend the notion of local Laplacians to the persistent setting. Let $\mathcal{K}_{\bullet} = \{\mathcal{K}_t\}_{t=0}^m$ be a filtration of simplicial complexes, and fix a vertex $v \in \mathcal{K}_0$. A filtration $\mathcal{K}_{\bullet}$ of simplicial complexes is defined as a nested sequence:
 \[
 \mathcal{K}_0 \subseteq \mathcal{K}_1 \subseteq \dots \subseteq \mathcal{K}_m.
 \]
 Then $v$ is contained in all subsequent complexes of the filtration by definition of a simplicial filtration. This gives rise to a filtration of relative simplicial complexes
 \[
 (\mathcal{K}_0, \mathcal{K}_0 \setminus \{v\}) \hookrightarrow (\mathcal{K}_1, \mathcal{K}_1 \setminus \{v\}) \hookrightarrow \dots \hookrightarrow (\mathcal{K}_m, \mathcal{K}_m \setminus \{v\}).
 \]
 For any indices $s \leq t$, the inclusion of the underlying simplicial complexes induces an inclusion of relative chain complexes
 \[
 \bar{j}_\ast: C_{\ast}(\mathcal{K}_s, \mathcal{K}_s \setminus \{v\}) \hookrightarrow C_{\ast}(\mathcal{K}_t, \mathcal{K}_t \setminus \{v\}).
 \]
 Recall that all relative chain complexes are equipped with the standard inner product induced by the basis of simplices. With respect to this inner product structure, the map $\bar{j}_\ast$ is an isometric embedding of differential graded inner product spaces.

 \begin{definition}[Persistent Local Laplacian]
 	For any $s \leq t$, the $k$-th \textit{$(s,t)$-persistent local Laplacian} $\Delta_k^{s,t}(\mathcal{K}_{\bullet}; v):C_{k}(\mathcal{K}_s, \mathcal{K}_s \setminus \{v\})
 	\to C_{k}(\mathcal{K}_s, \mathcal{K}_s \setminus \{v\})$ of the filtration $\mathcal{K}_{\bullet}$ at vertex $v$ is the operator
 	\begin{equation*}
 		\Delta_k^{s,t}(\mathcal{K}_{\bullet}; v) := (\overline{\partial}_k^s)^* \overline{\partial}_k^s + \overline{\partial}_{k+1}^{s,t} (\overline{\partial}_{k+1}^{s,t})^*, \quad k \geq 0.
 	\end{equation*}
 \end{definition}
 
 Analogous to the non-persistent case, the persistent local Laplacian $\Delta_k^{s,t}(\mathcal{K}_{\bullet}; v)$ is a self-adjoint, positive semi-definite operator. The kernel of this operator, the space of harmonic vectors with respect to $\Delta_k^{s,t}(\mathcal{K}_{\bullet}; v)$, is called the $(s,t)$-persistent local harmonic space, denoted $\mathcal{H}_k^{s,t}(\mathcal{K}_{\bullet}; v)$.

 \subsubsection{Persistent Link Laplacian}
 
 Recall from the previous subsection that for any simplicial complex $\mathcal{K}$ and a vertex $v \in \mathcal{K}$, the link complex $\Lk_{\mathcal{K}}(v)$ can be constructed, which serves as a key tool for computing the local Laplacian of $\mathcal{K}$ at $v$. Extending this construction to the persistent setting, we first derive a filtration of link complexes from a given simplicial filtration.
 
 Let $\mathcal{K}_{\bullet} = \{\mathcal{K}_t\}_{t=0}^m$ be a filtration of simplicial complexes, and fix a vertex $v \in \mathcal{K}_0$. For each $t$, the link complex $\Lk_{\mathcal{K}_t}(v)$ is a subcomplex of $\Lk_{\mathcal{K}_{t+1}}(v)$. This gives rise to a filtration of link complexes:
 \[
 \Lk_{\mathcal{K}_0}(v) \hookrightarrow \Lk_{\mathcal{K}_1}(v) \hookrightarrow \dots \hookrightarrow \Lk_{\mathcal{K}_m}(v).
 \]
 We denote this link filtration by $\Lk_{\mathcal{K}_{\bullet}}(v) = \{\Lk_{\mathcal{K}_t}(v)\}_{t=0}^m$. Since $\Lk_{\mathcal{K}_{\bullet}}(v)$ is a filtration of simplicial complexes, we can define the standard persistent Laplacian on this filtration, following the same construction as the persistent Laplacian. We denote this persistent Laplacian by $\Delta_k^{s,t}(\Lk_{\mathcal{K}_{\bullet}}(v))$, where $s \leq t$ and $k \geq 0$ denote the persistence interval and homological dimension, respectively.
 
 The core result is a fundamental relationship between the persistent local Laplacian and the persistent link Laplacian, which generalizes the non-persistent connection between local and link Laplacians.

 \begin{theorem}
 	Let $\mathcal{K}_{\bullet} = \{\mathcal{K}_t\}_{t=0}^m$ be a filtration of simplicial complexes, and fix a vertex $v \in \mathcal{K}_0$. For any indices $s \leq t$ and integer $k \geq 0$, the $(s,t)$-persistent local Laplacian $\Delta_k^{s,t}(\mathcal{K}_{\bullet}; v)$ is unitarily equivalent to the $(s,t)$-persistent Laplacian of the link filtration $\Lk_{\mathcal{K}_{\bullet}}(v)$ at dimension $k-1$, i.e.,
 	\[
 	\Delta_k^{s,t}(\mathcal{K}_{\bullet}; v) \cong \Delta_{k-1}^{s,t}(\Lk_{\mathcal{K}_{\bullet}}(v)).
 	\]
 	As a direct consequence, their kernels (persistent harmonic spaces) are isomorphic
 	\[
 	\ker \Delta_k^{s,t}(\mathcal{K}_{\bullet}; v) \cong \ker \Delta_{k-1}^{s,t}(\Lk_{\mathcal{K}_{\bullet}}(v)).
 	\]
 \end{theorem}
 
 This unitary equivalence implies that the persistent local Laplacian $\Delta_k^{s,t}(\mathcal{K}_{\bullet}; v)$ and the persistent link Laplacian $\Delta_{k-1}^{s,t}(\Lk_{\mathcal{K}_{\bullet}}(v))$ have identical spectral properties.

 
 Beyond its theoretical significance, the persistent link Laplacian also provides a more efficient mathematical approach for computing persistent local spectral features. Leveraging the unitary equivalence established in the above theorem, this approach enables the design of more efficient computer algorithms. In turn, these algorithms lay a solid mathematical foundation for potential large-scale data computation scenarios, addressing the computational challenges posed by massive datasets in persistent topology applications.

 \subsection{Persistent local Laplacian molecular descriptors}
 
 There are various types of intra- and intermolecular interactions, including hydrogen bonding, electrostatic forces, van der Waals forces, and hydrophobic/hydrophilic interactions. To effectively characterize these critical interactions, element-specific modeling was developed in our earlier work \cite{cang2017topologynet}, where it demonstrated strong predictive power. In this study, we integrate element-specific modeling with the proposed PLL framework to capture and encode these interactions into molecular features. Specifically, we partition atoms in a protein--ligand complex into element-based groups. The four most common atom types in proteins are \{C, N, O, S\}, while ligands frequently contain \{C, N, O, S, P, Cl, Br, I, H\}. This yields $40$ possible protein--ligand atom-type interaction pairs. A standard cutoff distance of 12~\AA{} from the ligand is used to collect neighboring protein atoms. For each resulting 3D point cloud associated with an atom-type pair, we apply PLL to generate a vectorized representation. The concatenation of the vectorized features from all 40 atom-type combinations forms the molecular descriptor for a protein--ligand complex. In biomolecular featurization, we consider the topological multi-scale features that are not the persistent local Laplacian model, but rather the local Laplacian corresponding to each filtration parameter $s$.
  
Specifically, for each protein--ligand atom-type combination, we perform PLL analysis centered at every atom $v$, using a Vietoris-Rips complex with a filtration radius ranging from 2 to 6~\AA{} in increments of 0.5~\AA{}. For a given filtration radius, we obtain the eigenvalues of the dimension-1 local Laplacian matrix $\Delta_1^{s,s}(\mathcal{K}_{\bullet}; v)$ (computed through link Laplacian $\Delta_0^{s,s}(\Lk_{\mathcal{K}_{\bullet}}(v))$) and summarize them using nine statistical descriptors, including the sum, mean, median, standard deviation, maximum, minimum, and variance, as well as the sum of squared positive eigenvalues and the number of zero eigenvalues. These nine values form the feature representation for each atom. To obtain a scalable representation for one atomic point cloud, we categorize atoms into two groups---protein atoms and ligand atoms. For each of the nine per-atom feature types, we compute two aggregate statistics (sum and mean) across all atoms in each group. This yields 18 features for protein atoms and 18 features for ligand atoms, giving a total of 36 features at a single filtration radius. Because we use nine filtration radii, we obtain 324 features per point-cloud combination. In total, 12{,}960 features are generated to characterize a protein--ligand complex. We also construct ligand-specific molecular descriptors using PLL analysis. For each ligand, we consider the point cloud formed by atoms of types C, N, O, S, P, Cl, Br, and I, and apply a Vietoris-Rips filtration from 2 to 6~\AA{} with a step size of 0.5~\AA{}. We further perform element-specific characterization using 13 atom subsets: \{C\}, \{N\}, \{O\}, \{S\}, \{C, N\}, \{C, O\}, \{C, S\}, \{N, O\}, \{N, S\}, \{O, S\}, \{N, P\}, \{F, Cl, Br, I\}, and \{C, O, N, S, F, P, Cl, Br, I\}. For each subset, we obtain the same nine statistical descriptors of the dimension-1 local Laplacian $\Delta_1^{s,s}(\mathcal{K}_{\bullet}; v)$ eigenvalues at each filtration radius, and then take the sum and mean of each descriptor across all atoms in the subset. This procedure yields 2{,}106 features for each ligand. Finally, we concatenate the 12{,}960 protein--ligand complex features with the 2{,}106 ligand features, producing a representation that captures both intermolecular and intramolecular interactions for each protein--ligand pair.

 \subsection{Natural language processing (NLP) molecular descriptors}

Transformer models, a core class of natural language processing (NLP) techniques, have recently emerged as powerful tools in molecular biosciences. In this work, we employ pretrained Transformer models to extract sequence-based features for both proteins \cite{rives2021biological} and ligands \cite{chen2021extracting}. The inputs to these models are amino acid sequences for proteins and SMILES strings for ligands. By concatenating the molecular descriptors generated from the pretrained models (ESM) \cite{rives2021biological} for protein sequences and ligand SMILES strings \cite{chen2021extracting}, we obtain a unified sequence-based representation of each protein–ligand complex. The details of these two methods are provided in Section 5 in the Supporting Information.

\subsection{Machine learning modeling}

To construct our predictive models, we adopt the Gradient Boosting Decision Tree (GBDT) framework, implemented through the \texttt{scikit-learn} library (v1.5.1). GBDT offers strong performance in a wide range of regression tasks due to its resilience to overfitting, minimal sensitivity to hyperparameter tuning, and straightforward training procedure. The method iteratively builds an ensemble of decision trees, where each tree is trained to correct the residual errors of the previous ones. Although individual trees act as weak learners, their sequential combination forms a powerful model capable of capturing complex nonlinear relationships. Both the PLL-derived structural features and Transformer-generated sequence features are independently used as inputs to train GBDT regressors. The specific hyperparameter settings employed in our experiments are summarized in \autoref{table:GBDT-parameters}. The formulations of the two evaluation metrics including PCC and RMSE are given in Section 6 in the Supporting Information.

\begin{table}[htb!]
	\small
	\centering
	\begin{tabular}{c c c  c }
		\hline
		No. of estimators &  Max depth & Min. sample split & Learning rate\\
		30000/20000& 7 &5 & 0.002\\
		\hline
		Max features & Subsample size & Repetition &\\
		Square root & 0.8&  10 times & \\
		\hline
	\end{tabular}
	\caption{Hyperparameters used for build gradient boosting regression models. Tree numbers are set to be 30000 and 20000 respectively for PLL and transformer-based biomolecular modeling.}
	\label{table:GBDT-parameters}
\end{table}

\section{Discussion}


\begin{figure}[!htb]
	\centering
	\includegraphics[width=0.82\textwidth]{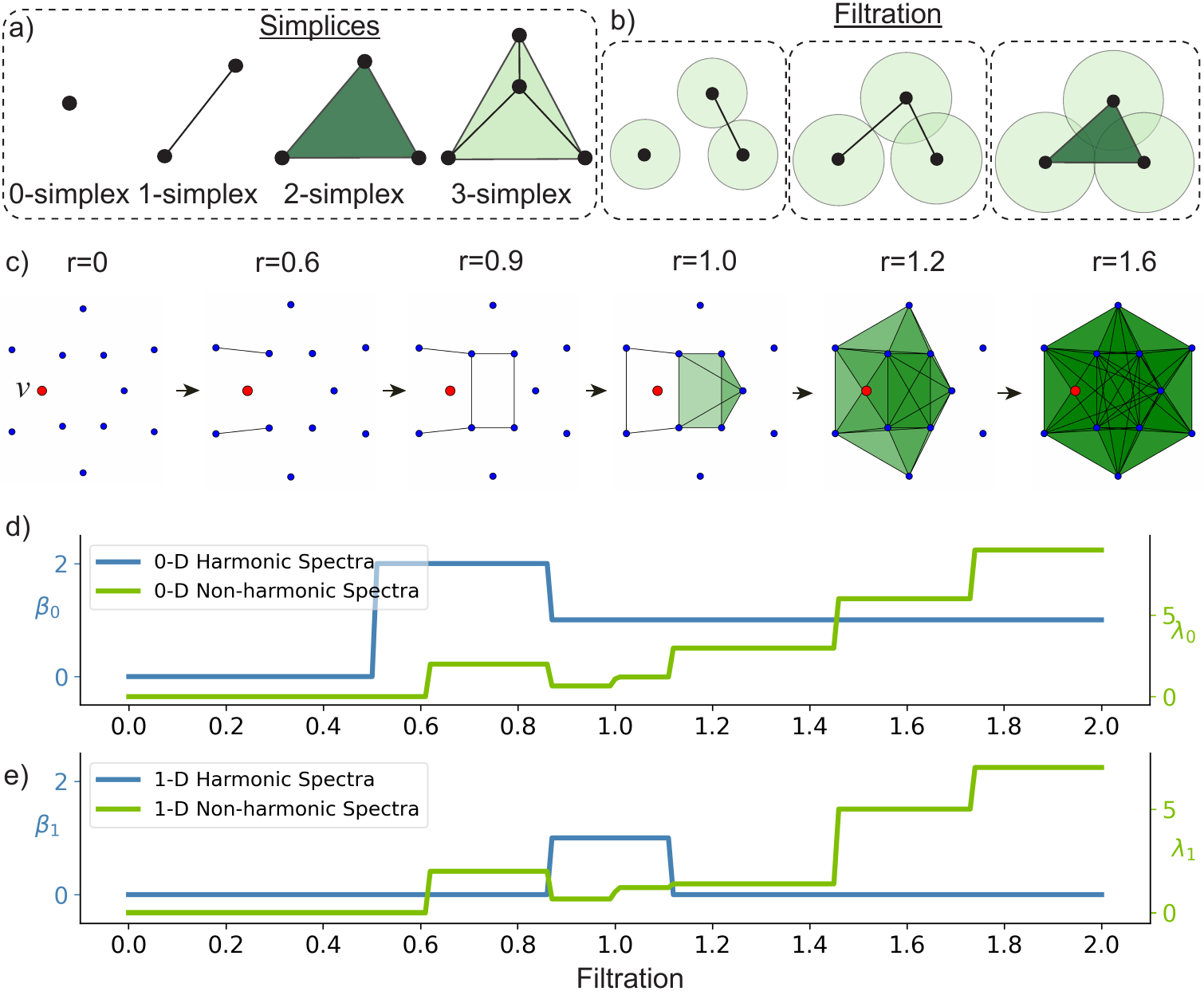}
	\caption{ Illustration of persistent local Laplacians (PLL) at one point in a point cloud. (c): a filtration process of the Vietoris-Rips complex constructed from a point cloud of 12 points. Among the 12 points, six lie on a circle of radius $1$ and the other six lie on a circle of radius $0.5$. The points on the smaller circle are obtained by a rotation of $\frac{\pi}{6}$ so that each point on the smaller circle lies on the same radial line from the origin as a corresponding point on the larger circle. (d): Analysis of zero-dimensional spectral information of local Laplacian matrices defined for the red point $v$ over the filtration process. (e): Analysis of one-dimensional persistent Laplacian spectra of PLL defined for point $v$ over the filtration process.}
	\label{Fig:demon-PLL-v3}
\end{figure}

\subsection{PLL Modeling interpretability}

 PLL framework offers excellent interpretability for modeling molecular structures, which motivates its application to molecular descriptor design. To illustrate the PLL modeling process, we consider a point-cloud example shown in \autoref{Fig:demon-PLL-v3}. \autoref{Fig:demon-PLL-v3}a show 4 types of building blocks in a simplicial complex, while panel b illustrates a filtration process using Vietoris--Rips complexes. In \autoref{Fig:demon-PLL-v3}c, a Vietoris--Rips filtration is applied to a point cloud consisting of 12 points, where six points lie on a circle of radius $1$ and the remaining six lie on a circle of radius $0.5$. Link complex $\Lk_{\mathcal{K}_{\bullet}}(v)$ encodes the local topological and geometric structures in the neighborhood of a point of interest ($v$, highlighted in red). It is composed of edges and triangles around $v$. As the filtration radius increases, link complexes of progressively larger sizes are formed around $v$. 

For each link complex, local Laplacian matrices of different dimensions can be constructed, and their spectral properties quantitatively characterize the local structural features surrounding $v$. The Betti number $\beta_0$, defined as the number of zero eigenvalues of the link Laplacian matrix $\Delta_{0}^{s,s}(\Lk_{\mathcal{K}_{\bullet}}(v))$, indicates the number of independent connected components in the link complex. We plot the variation of $\beta_0$ along the filtration as the blue curve in \autoref{Fig:demon-PLL-v3}d. At filtration radius $r = 0.0$, the link complex of the red point is empty, yielding $\beta_0 = 0$ in \autoref{Fig:demon-PLL-v3}d. When the filtration radius increases to $r = 0.6$, two independent edges appear in the link complex, resulting in $\beta_0 = 2$. As the filtration radius further increases to $r = 1.0$, these edges become connected, forming a single connected component and reducing $\beta_0$ to $1$. Beyond the zero eigenvalues, the nonzero eigenvalues of local Laplacian matrices provide additional geometric information and reveal the homotopic evolution of the link sets throughout the filtration. In particular, the smallest nonzero eigenvalues $\Lambda_0$ reflect geometric changes in the local structures as the filtration progresses, as shown by the green curve in \autoref{Fig:demon-PLL-v3}d. The values of $\Lambda_0$ capture connectivity changes in these link complexes.

Similarly, the Betti number $\beta_1$, defined as the number of zero eigenvalues of the link Laplacian matrix $\Delta_{1}^{s,s}(\Lk_{\mathcal{K}_{\bullet}}(v))$, quantifies the number of independent loops in the link complex. We compute these values along the filtration, as shown by the blue curve in \autoref{Fig:demon-PLL-v3}e. At a filtration radius of $r = 1.0$, a single loop appears in the link complex (see the third figure in \autoref{Fig:demon-PLL-v3}c), giving $\beta_1 = 1$. These spectral quantities derived from the link Laplacians $\Delta_{k}^{s,s}(\Lk_{\mathcal{K}_{\bullet}}(v))$ provide both topological and geometric insights into local structures, making them well suited for biomolecular structural modeling.

\subsection{Filtration impacts on modeling performance}

The performance of PLL in predicting protein–ligand binding depends on the parameterization of the filtration. To determine the optimal filtration range, we evaluated model performance under different filtration radii on the three PDBbind benchmark datasets. As shown in Figure 3 in the Supporting Information, the predictive performance of the PLL model improves as the filtration radius increases from 4\AA~to 6\AA. However, performance declines slightly beyond 6\AA~on PDBbind-v2007 and PDBbind-v2016, whereas on PDBbind-v2013 the model achieves its best performance at 7\AA, with only marginal improvement over 6\AA. These results suggest that a filtration radius of 6\AA~is optimal for PLL-based modeling of protein–ligand binding. Using an upper filtration radius of 6\AA~also enables efficient feature construction while maintaining the ability to capture essential molecular interactions.


\subsection{Computational complexity analysis}

The computational complexity of the proposed persistent link Laplacian is significantly lower than that of the standard persistent Laplacian method. In both approaches, the computational cost primarily arises from two sources: (i) the construction of the Laplacian matrix and (ii) the computation of its full spectrum of eigenvalues. The persistent link Laplacian exhibits lower complexity in both aspects. Detailed analyses and comparisons of the computational complexity are provided in subsection 3.3 and Table S1 of the Supporting Information.

 \section*{Data Availability}
  
  All data and the code needed to reproduce this paper's result can be found at\\ \href{https://github.com/hfeng2uccc/PLLML.git}{https://github.com/hfeng2uccc/PLLML.git}. The PDBbind datasets are available at \url{https://www.pdbbind-plus.org.cn/}.

\section*{Acknowledgments}
This work was supported in part by Natural Science Foundation of China (NSFC Grant No.12401080), Scientific Research Foundation of Chongqing University of Technology, and Start-up Research Fund of University of North Carolina at Charlotte.



\begin{thebibliography}{10}
	
	\bibitem{fleming2018computer}
	Nic Fleming.
	\newblock Computer-calculated compounds.
	\newblock {\em Nature}, 557(7707):S55--7, 2018.
	
	\bibitem{lyu2019ultra}
	Jiankun Lyu, Sheng Wang, Trent~E Balius, Isha Singh, Anat Levit, Yurii~S Moroz,
	Matthew~J O'Meara, Tao Che, Enkhjargal Algaa, Kateryna Tolmachova, et~al.
	\newblock Ultra-large library docking for discovering new chemotypes.
	\newblock {\em Nature}, 566(7743):224--229, 2019.
	
	\bibitem{kitchen2004docking}
	Douglas~B Kitchen, H{\'e}l{\`e}ne Decornez, John~R Furr, and J{\"u}rgen
	Bajorath.
	\newblock Docking and scoring in virtual screening for drug discovery: methods
	and applications.
	\newblock {\em Nature reviews Drug discovery}, 3(11):935--949, 2004.
	
	\bibitem{pinzi2019molecular}
	Luca Pinzi and Giulio Rastelli.
	\newblock Molecular docking: shifting paradigms in drug discovery.
	\newblock {\em International journal of molecular sciences}, 20(18):4331, 2019.
	
	\bibitem{pagadala2017software}
	Nataraj~S Pagadala, Khajamohiddin Syed, and Jack Tuszynski.
	\newblock Software for molecular docking: a review.
	\newblock {\em Biophysical reviews}, 9:91--102, 2017.
	
	\bibitem{wang2015accurate}
	Lingle Wang, Yujie Wu, Yuqing Deng, Byungchan Kim, Levi Pierce, Goran Krilov,
	Dmitry Lupyan, Shaughnessy Robinson, Markus~K Dahlgren, Jeremy Greenwood,
	et~al.
	\newblock Accurate and reliable prediction of relative ligand binding potency
	in prospective drug discovery by way of a modern free-energy calculation
	protocol and force field.
	\newblock {\em Journal of the American Chemical Society}, 137(7):2695--2703,
	2015.
	
	\bibitem{sliwoski2014computational}
	Gregory Sliwoski, Sandeepkumar Kothiwale, Jens Meiler, and Edward~W Lowe.
	\newblock Computational methods in drug discovery.
	\newblock {\em Pharmacological reviews}, 66(1):334--395, 2014.
	
	\bibitem{jumper2021highly}
	John Jumper, Richard Evans, Alexander Pritzel, Tim Green, Michael Figurnov,
	Olaf Ronneberger, Kathryn Tunyasuvunakool, Russ Bates, Augustin
	{\v{Z}}{\'\i}dek, Anna Potapenko, et~al.
	\newblock Highly accurate protein structure prediction with alphafold.
	\newblock {\em Nature}, 596(7873):583--589, 2021.
	
	\bibitem{baek2021accurate}
	Minkyung Baek, Frank DiMaio, Ivan Anishchenko, Justas Dauparas, Sergey
	Ovchinnikov, Gyu~Rie Lee, Jue Wang, Qian Cong, Lisa~N Kinch, R~Dustin
	Schaeffer, et~al.
	\newblock Accurate prediction of protein structures and interactions using a
	three-track neural network.
	\newblock {\em Science}, 373(6557):871--876, 2021.
	
	\bibitem{lin2023evolutionary}
	Zeming Lin, Halil Akin, Roshan Rao, Brian Hie, Zhongkai Zhu, Wenting Lu, Nikita
	Smetanin, Robert Verkuil, Ori Kabeli, Yaniv Shmueli, et~al.
	\newblock Evolutionary-scale prediction of atomic-level protein structure with
	a language model.
	\newblock {\em Science}, 379(6637):1123--1130, 2023.
	
	\bibitem{song2024multiobjective}
	Yao Song and Lu~Wang.
	\newblock Multiobjective tree-based reinforcement learning for estimating
	tolerant dynamic treatment regimes.
	\newblock {\em Biometrics}, 80(1):ujad017, 2024.
	
	\bibitem{li2015improving}
	Hongjian Li, Kwong-Sak Leung, Man-Hon Wong, and Pedro~J Ballester.
	\newblock Improving {AutoDock Vina} using random forest: the growing accuracy
	of binding affinity prediction by the effective exploitation of larger data
	sets.
	\newblock {\em Molecular informatics}, 34(2-3):115--126, 2015.
	
	\bibitem{feinberg2018potentialnet}
	Evan~N Feinberg, Debnil Sur, Zhenqin Wu, Brooke~E Husic, Huanghao Mai, Yang Li,
	Saisai Sun, Jianyi Yang, Bharath Ramsundar, and Vijay~S Pande.
	\newblock Potential{Net} for molecular property prediction.
	\newblock {\em ACS central science}, 4(11):1520--1530, 2018.
	
	\bibitem{ballester2010machine}
	Pedro~J Ballester and John~BO Mitchell.
	\newblock A machine learning approach to predicting protein--ligand binding
	affinity with applications to molecular docking.
	\newblock {\em Bioinformatics}, 26(9):1169--1175, 2010.
	
	\bibitem{wang2017improving}
	Cheng Wang and Yingkai Zhang.
	\newblock Improving scoring-docking-screening powers of protein--ligand scoring
	functions using random forest.
	\newblock {\em Journal of computational chemistry}, 38(3):169--177, 2017.
	
	\bibitem{pan2022aa}
	Xiaolin Pan, Hao Wang, Yueqing Zhang, Xingyu Wang, Cuiyu Li, Changge Ji, and
	John~ZH Zhang.
	\newblock Aa-score: a new scoring function based on amino acid-specific
	interaction for molecular docking.
	\newblock {\em Journal of Chemical Information and Modeling},
	62(10):2499--2509, 2022.
	
	\bibitem{nguyen2021graphdta}
	Thin Nguyen, Hang Le, Thomas~P Quinn, Tri Nguyen, Thuc~Duy Le, and Svetha
	Venkatesh.
	\newblock G{raphDTA}: predicting drug--target binding affinity with graph
	neural networks.
	\newblock {\em Bioinformatics}, 37(8):1140--1147, 2021.
	
	\bibitem{wallach2015atomnet}
	Izhar Wallach, Michael Dzamba, and Abraham Heifets.
	\newblock Atomnet: a deep convolutional neural network for bioactivity
	prediction in structure-based drug discovery.
	\newblock {\em arXiv preprint arXiv:1510.02855}, 2015.
	
	\bibitem{ragoza2017protein}
	Matthew Ragoza, Joshua Hochuli, Elisa Idrobo, Jocelyn Sunseri, and David~Ryan
	Koes.
	\newblock Protein--ligand scoring with convolutional neural networks.
	\newblock {\em Journal of chemical information and modeling}, 57(4):942--957,
	2017.
	
	\bibitem{nguyen2020review}
	Duc~Duy Nguyen, Zixuan Cang, and Guo-Wei Wei.
	\newblock A review of mathematical representations of biomolecular data.
	\newblock {\em Physical Chemistry Chemical Physics}, 22(8):4343--4367, 2020.
	
	\bibitem{meng2021persistent}
	Zhenyu Meng and Kelin Xia.
	\newblock Persistent spectral--based machine learning (perspect ml) for
	protein-ligand binding affinity prediction.
	\newblock {\em Science advances}, 7(19):eabc5329, 2021.
	
	\bibitem{liu2023persistent}
	Ran Liu, Xiang Liu, and Jie Wu.
	\newblock Persistent path-spectral (pps) based machine learning for
	protein--ligand binding affinity prediction.
	\newblock {\em Journal of Chemical Information and Modeling}, 63(3):1066--1075,
	2023.
	
	\bibitem{wang2020persistent}
	Rui Wang, Duc~Duy Nguyen, and Guo-Wei Wei.
	\newblock Persistent spectral graph.
	\newblock {\em International Journal for Numerical Methods in Biomedical
		Engineering}, 36(9):e3376, 2020.
	
	\bibitem{carlsson2009topology}
	Gunnar Carlsson.
	\newblock Topology and data.
	\newblock {\em Bulletin of the American Mathematical Society}, 46(2):255--308,
	2009.
	
	\bibitem{edelsbrunner2008persistent}
	Herbert Edelsbrunner and John Harer.
	\newblock Persistent homology---a survey.
	\newblock In {\em Surveys on Discrete and Computational Geometry: Twenty Years
		Later}, volume 453 of {\em Contemporary Mathematics}, pages 257--282.
	American Mathematical Society, 2008.
	
	\bibitem{nguyen2019mathematical}
	Duc~Duy Nguyen, Zixuan Cang, Kedi Wu, Menglun Wang, Yin Cao, and Guo-Wei Wei.
	\newblock Mathematical deep learning for pose and binding affinity prediction
	and ranking in d3r grand challenges.
	\newblock {\em Journal of computer-aided molecular design}, 33:71--82, 2019.
	
	\bibitem{nguyen2020mathdl}
	Duc~Duy Nguyen, Kaifu Gao, Menglun Wang, and Guo-Wei Wei.
	\newblock Mathdl: mathematical deep learning for d3r grand challenge 4.
	\newblock {\em Journal of computer-aided molecular design}, 34:131--147, 2020.
	
	\bibitem{memoli2022persistent}
	Facundo M{\'e}moli, Zhengchao Wan, and Yusu Wang.
	\newblock Persistent laplacians: Properties, algorithms and implications.
	\newblock {\em SIAM Journal on Mathematics of Data Science}, 4(2):858--884,
	2022.
	
	\bibitem{chen2024multiscale}
	Dong Chen, Jian Liu, and Guo-Wei Wei.
	\newblock Multiscale topology-enabled structure-to-sequence transformer for
	protein--ligand interaction predictions.
	\newblock {\em Nature Machine Intelligence}, 6(7):799--810, 2024.
	
	\bibitem{qiu2023persistent}
	Yuchi Qiu and Guo-Wei Wei.
	\newblock Persistent spectral theory-guided protein engineering.
	\newblock {\em Nature Computational Science}, 3(2):149--163, 2023.
	
	\bibitem{chen2022persistent}
	Jiahui Chen, Yuchi Qiu, Rui Wang, and Guo-Wei Wei.
	\newblock Persistent laplacian projected omicron ba. 4 and ba. 5 to become new
	dominating variants.
	\newblock {\em Computers in Biology and Medicine}, 151:106262, 2022.
	
	\bibitem{liu2026local}
	Jian Liu, Hongsong Feng, and Kefeng Liu.
	\newblock Local laplacian: theory and models for data analysis.
	\newblock {\em arXiv preprint arXiv:2603.07591}, 2026.
	
	\bibitem{hatcher2002algebraic}
	Allen Hatcher.
	\newblock {\em Algebraic Topology}.
	\newblock Cambridge University Press, 2002.
	
	\bibitem{liu2015pdb}
	Zhihai Liu, Yan Li, Li~Han, Jie Li, Jie Liu, Zhixiong Zhao, Wei Nie, Yuchen
	Liu, and Renxiao Wang.
	\newblock Pdb-wide collection of binding data: current status of the pdbbind
	database.
	\newblock {\em Bioinformatics}, 31(3):405--412, 2015.
	
	\bibitem{feng2025mayer}
	Hongsong Feng, Li~Shen, Jian Liu, and Guo-Wei Wei.
	\newblock Mayer-homology learning prediction of protein-ligand binding
	affinities.
	\newblock {\em Journal of computational biophysics and chemistry},
	24(02):253--266, 2025.
	
	\bibitem{nguyen2019agl}
	Duc~Duy Nguyen and Guo-Wei Wei.
	\newblock {AGL}-score: algebraic graph learning score for protein--ligand
	binding scoring, ranking, docking, and screening.
	\newblock {\em Journal of chemical information and modeling}, 59(7):3291--3304,
	2019.
	
	\bibitem{shen2024knot}
	Li~Shen, Hongsong Feng, Fengling Li, Fengchun Lei, Jie Wu, and Guo-Wei Wei.
	\newblock Knot data analysis using multiscale gauss link integral.
	\newblock {\em Proceedings of the National Academy of Sciences},
	121(42):e2408431121, 2024.
	
	\bibitem{cang2017topologynet}
	Zixuan Cang and Guo-Wei Wei.
	\newblock {TopologyNet}: Topology based deep convolutional and multi-task
	neural networks for biomolecular property predictions.
	\newblock {\em PLoS Computational Biology}, 13(7):e1005690, 2017.
	
	\bibitem{feng2025caml}
	Hongsong Feng, Faisal Suwayyid, Mushal Zia, JunJie Wee, Yuta Hozumi, Chun-Long
	Chen, and Guo-Wei Wei.
	\newblock Caml: Commutative algebra machine learning--a case study on
	protein--ligand binding affinity prediction.
	\newblock {\em Journal of Chemical Information and Modeling}, 2025.
	
	\bibitem{rives2021biological}
	Alexander Rives, Joshua Meier, Tom Sercu, Siddharth Goyal, Zeming Lin, Jason
	Liu, Demi Guo, Myle Ott, C~Lawrence Zitnick, Jerry Ma, et~al.
	\newblock Biological structure and function emerge from scaling unsupervised
	learning to 250 million protein sequences.
	\newblock {\em Proceedings of the National Academy of Sciences},
	118(15):e2016239118, 2021.
	
	\bibitem{chen2021extracting}
	Dong Chen, Jiaxin Zheng, Guo-Wei Wei, and Feng Pan.
	\newblock Extracting predictive representations from hundreds of millions of
	molecules.
	\newblock {\em The journal of physical chemistry letters}, 12(44):10793--10801,
	2021.
	
	\bibitem{li2023development}
	Chuang Li, Aiwei Zhang, Lifei Wang, Jiaqi Zuo, Caizhen Zhu, Jian Xu, Mingliang
	Wang, and John~ZH Zhang.
	\newblock Development of a polynomial scoring function p3-score for improved
	scoring and ranking powers.
	\newblock {\em Chemical Physics Letters}, 824:140547, 2023.
	
	\bibitem{cang2018representability}
	Zixuan Cang, Lin Mu, and Guo-Wei Wei.
	\newblock Representability of algebraic topology for biomolecules in machine
	learning based scoring and virtual screening.
	\newblock {\em PLoS computational biology}, 14(1):e1005929, 2018.
	
	\bibitem{cheng2009comparative}
	Tiejun Cheng, Xun Li, Yan Li, Zhihai Liu, and Renxiao Wang.
	\newblock Comparative assessment of scoring functions on a diverse test set.
	\newblock {\em Journal of chemical information and modeling}, 49(4):1079--1093,
	2009.
	
	\bibitem{li2014comparative}
	Yan Li, Zhihai Liu, Jie Li, Li~Han, Jie Liu, Zhixiong Zhao, and Renxiao Wang.
	\newblock Comparative assessment of scoring functions on an updated benchmark:
	1. compilation of the test set.
	\newblock {\em Journal of chemical information and modeling}, 54(6):1700--1716,
	2014.
	
	\bibitem{su2018comparative}
	Minyi Su, Qifan Yang, Yu~Du, Guoqin Feng, Zhihai Liu, Yan Li, and Renxiao Wang.
	\newblock Comparative assessment of scoring functions: the casf-2016 update.
	\newblock {\em Journal of chemical information and modeling}, 59(2):895--913,
	2018.
	
\end{thebibliography}

\end{document}